\documentclass[twocolumn,multicol]{aastex62}

\usepackage{amsmath,amstext,amssymb}
\usepackage{apjfonts}
\usepackage{comment}
\usepackage{footnote}
\usepackage{soul}
\usepackage{booktabs} 
\usepackage{multirow} 
\usepackage{threeparttablex, tablefootnote}
\usepackage[figure,figure*]{hypcap}

\shorttitle{\textit{NICER} and \textit{XMM} observation of RX J1605.3+3249}
\shortauthors{Malacaria et al. 2019}

\begin{document}
\title{A joint \textit{NICER} and \textit{XMM-Newton} view of the ``Magnificent'' \\ thermally emitting X-ray Isolated Neutron Star RX J1605.3+3249}

\author[0000-0002-0380-0041]{Christian Malacaria}
\affiliation{NASA Marshall Space Flight Center, NSSTC, 320 Sparkman Drive, Huntsville, AL 35805, USA}\thanks{NASA Postdoctoral Fellow}
\affiliation{Universities Space Research Association, NSSTC, 320 Sparkman Drive, Huntsville, AL 35805, USA}
\author[0000-0002-9870-2742]{Slavko Bogdanov}
\affiliation{Columbia Astrophysics Laboratory, Columbia University, 550 West 120th Street, New York, NY 10027}
\author[0000-0002-6089-6836]{Wynn C. G. Ho}
\affiliation{Department of Physics and Astronomy, Haverford College, 370 Lancaster Avenue, Haverford, PA 19041, USA}
\affiliation{Mathematical Sciences, Physics and Astronomy, and STAG Research Centre, University of Southampton, Southampton SO17 1BJ, United Kingdom}
\author[0000-0003-1244-3100]{Teruaki Enoto}
\affiliation{Department of Astronomy, Kyoto University, Kitashirakawa-Oiwake-cho, Sakyo-ku, Kyoto 606-8502, Japan}
\author[0000-0002-5297-5278]{Paul S. Ray}
\affiliation{Space Science Division, U.S. Naval Research Laboratory, Washington, DC 20375-5352, USA}
\author{Zaven Arzoumanian}
\affiliation{X-ray Astrophysics Laboratory, Astrophysics Science Division, NASA's Goddard Space Flight Center, Greenbelt, MD 20771, USA}
\author{Thoniel Cazeau}
\affiliation{X-ray Astrophysics Laboratory, Astrophysics Science Division, NASA's Goddard Space Flight Center, Greenbelt, MD 20771, USA}
\author{Keith C. Gendreau}
\affiliation{X-ray Astrophysics Laboratory, Astrophysics Science Division, NASA's Goddard Space Flight Center, Greenbelt, MD 20771, USA}
\author[0000-0002-6449-106X]{Sebastien Guillot} \affil{CNRS, IRAP, 9 avenue du Colonel Roche, BP 44346, F-31028 Toulouse Cedex 4, France} ?
\author[0000-0002-3531-9842]{Tolga G{\"u}ver} \affiliation{Department of Astronomy and Space Sciences, Science Faculty, Istanbul University, Beyaz\i t, 34119 Istanbul, Turkey} \affiliation{Istanbul University Observatory Research and Application Center, Beyaz\i t, 34119 Istanbul, Turkey}
\author[0000-0002-6789-2723]{Gaurava K. Jaisawal} \affil{National Space Institute, Technical University of Denmark, Elektrovej, DK-2800 Lyngby, Denmark}
\author[0000-0002-4013-5650]{Michael T. Wolff}
\affil{Space Science Division, U.S. Naval Research Laboratory, Washington, DC 20375-5352, USA}
\collaboration{on behalf of the NICER Magnetars \& Magnetospheres Team}
\noaffiliation

  \begin{@twocolumnfalse}
\begin{abstract}

Thermally emitting X-ray isolated neutron stars represent excellent targets for testing cooling surface emission and atmosphere models, which are used to infer physical parameters of the neutron star. 
Among the seven known members of this class, RX~J1605.3+3249 is the only one that still lacks confirmation of its spin period.
Here we analyze \textit{NICER} and \textit{XMM-Newton} observations of RX~J1605.3+3249, in order to address its timing and spectral behavior.
Contrary to a previous tentative detection, but in agreement
with the recent work by \citet{Pires19}, we find no significant pulsation with pulsed fraction higher than $1.3\%$ ($3\sigma$) for periods above 150 ms. We also find a limit of $2.6\%$ for periods above 2 ms, despite searches in different energy bands.
The X-ray spectrum can be fit by either a double-blackbody model or by a single-temperature magnetized atmosphere model, both modified by a Gaussian absorption line at $\sim0.44\,$keV.
The origin of the absorption feature as a proton cyclotron line or as an atomic transition in the neutron star atmosphere is discussed.
The predictions of the best-fit X-ray models extended to  IR, optical and UV bands are compared with archival data.
Our results are interpreted in the framework of a fallback disk scenario.

    \end{abstract}
  \end{@twocolumnfalse}

\section{Introduction}\label{sec:intro}

Isolated Neutron Stars (INSs) are a class of neutron stars (NSs) with no evidence for any stellar companion and an undetected or extremely weak radio counterpart \citep{Kaspi06}.
Among these is the group of thermally emitting INSs discovered by the ROSAT satellite, seven in number, and hence dubbed the ``Magnificent Seven'' (M7, \citealt{Haberl07,Kaplan08,2009ASSL..357..141T}), otherwise called thermally emitting X-ray Isolated Neutron Stars (XINS, \citealt{Potekhin+Luca15}).

XINSs are thought to be members of the nearby OB associations of the Gould Belt \citep{Walter01,Popov03,Motch03,Motch05,Motch09}, located within a few hundreds of pc from the Sun \citep{Posselt07,Kerkwijk07}.
Their relatively low X-ray luminosity ($\sim10^{31-32}\,$erg\,s$^{-1}$) is consistent with cooling NSs of age $\sim0.5\,$Myr \citep{Page04}, in rough agreement with estimates from their kinematic ages \citep{Page06, Haberl07,Motch09}.
Their X-ray spectra are soft, thermal in nature, and usually fitted by blackbody components with temperatures in the range kT$\sim40-100\,$eV (e.g., \citealt{Kerkwijk07,Haberl07}).
Atmospheric emission models have also been proposed to fit the thermal emission from INSs (see \citealt{Potekhin16} and references therein). However, low-magnetic-field atmosphere models \citep{Gansicke02,Zavlin02} fail to reproduce the spectra of M7 members.
On the other hand, for the highly-magnetized atmosphere models, high-metallicity (e.g., iron) models predict a large number of absorption lines that are not observed, while pure hydrogen models reproduce spectral continua that are similar overall to the observed ones, \textit{and} imply an effective temperature considerably lower than that suggested by blackbody models \citep{Pavlov96, Ho+07}.
Finally, while blackbody models underachieve the observed flux at optical wavelengths \citep{Pavlov96,Burwitz03}, atmospheric models overestimate it in some cases.

Timing studies of the M7 members in X-rays reveal spin periods in the range $3-17\,$s and relatively high magnetic fields, B${\sim}10^{13}\,$G \citep{Kaplan11,Haberl07,Hambaryan17}.
Further commonality is observed among most M7 members, whose X-ray spectra show broad absorption features at energies $\sim0.2-0.8\,$keV  \citep{Haberl03,Zane05,Haberl07,Kerkwijk07,Schwope07}.
Such features are usually attributed either to electron/proton cyclotron resonant scattering features (eCRSF), proton cyclotron absorption features (pCF), or to bound-bound/bound-free transitions in atoms of strongly magnetized NS hydrogen atmospheres \citep{Zavlin02,Kerkwijk07}.

The only member of the M7 group that still lacks a coherent timing solution is RX~J1605.3+3249 (J1605 hereafter, \citealt{Schwope99}).
A possible candidate spin period of $6.9\,$s was tentatively proposed by \citet{Haberl07} but not confirmed in later observations.
Another possible candidate spin period of $3.4\,$s (and a spin period derivative of $\dot{P}\sim1.6\times10^{-12}\,$s\,s$^{-1}$) was proposed for J1605 by \citet{Pires+14}, which however was significant only at a low confidence level ($\sim4\sigma$).

The source distance has been analyzed in several works.
\citet{Posselt07} find two solutions for the distance, namely $390$ or $325\,$pc. 
On the other hand, \citet{Motch99} consider closer distance values, as low as $\sim100\,$pc, while \citet{Motch05} link the source with the Sco OB2 association within the Gould Belt, at a mean distance of $120-140\,$pc.
\citet{2012PASA...29...98T} argue that J1605 was probably born in the Octans association from a supernova at $\approx 100\,$pc and calculates the current distance of the NS as $300-400\,$pc.

A spectral feature at $\sim0.45\,$keV was first discovered by \citet{Kerkwijk04}.
\citet{Haberl07} found two additional absorption lines in the spectrum of J1605 obtained with \textit{XMM-Newton}, at energies of $0.59$ and $0.78\,$keV, and consistent with energies in a $2{:}3{:}4$ ratio.
A narrow absorption feature at $0.58\,$keV is also found by \citet{Kerkwijk04} and \citet{Hohle12} using the high energy-resolution instrument (RGS) on-board \textit{XMM-Newton}, with a width of $3.3\,$eV.
Further analysis of \textit{XMM-Newton} data by \citet{Pires+14} also finds significant absorption lines in J1605 at slightly different energies than previous works, that is $0.44$, $0.58$, and $0.83\,$keV, while \citet{Pires19} find no evidence of other absorption features than that at $\sim0.4\,$eV.

In this work we report the results of recent observations of J1605 performed with \textit{XMM-Newton} (\textit{XMM}) plus unpublished observations performed with the \textit{Neutron Star Interior Composition Explorer} (\textit{NICER}). 
We perform timing and spectral analysis in order to address open questions about J1605's main properties.
Combining both \textit{NICER} and \textit{XMM}, our observations do not show evidence of pulsations in the X-ray light curve of J1605.
Moreover, our long \textit{NICER} exposures allow to perform the most sensitive pulse search down to 2 ms to date.
The measured spectra are well fitted by a double-blackbody model or by a magnetized hydrogen atmospheric emission model, and we also confirm the presence of a broad absorption feature at $\sim0.45\,$keV.
Our timing and spectral results are consistent with the most recent work by \citet{Pires19}.
We integrate archival IR/Optical/UV data in our study, and interpret our findings in the context of the most recent XINS emission scenarios and the incidence of emission and absorption of their surrounding medium.

\section{Observations and data reduction}

In this work, we analyzed available \textit{NICER} data for J1605, and  only the most recent \textit{XMM} data. The \textit{NICER} and \textit{XMM} observation log is reported in Table~\ref{table:log}.

\subsection{NICER}

The \textit{NICER} X-ray Timing Instrument (XTI, \citealt{Gendreau16}) is an array of $56$ co-aligned X-ray concentrator optics, each associated with a silicon drift detector sensitive in the $0.2-12\,$keV band \citep{Prigozhin12}.
A single concentrator consists of $24$ nested  grazing-incidence gold-coated aluminum foil mirrors, parabolically shaped with a common focal length.
To date, $52$ detectors are operating, providing a peak effective area of $1900\,$cm$^2$ at $\sim1.5\,$keV, with an energy resolution of $\sim100\,$eV and a photon time-tagging resolution of $\sim100\,$ns.
With an effective area at 1 keV of about $2000\,$cm$^2$ (that is about 2 times that of \textit{XMM}-pn and several times that of \textit{Chandra}/ACIS), \textit{NICER} perfectly matches the needs to analyze soft, thermal emission from NSs.

\textit{NICER} observed J1605 for a total of $32$ segments between 2017 July 19 (Obs ID: 1032020101) and 2018 April 8 (Obs ID: 103202032), collecting a total of $\sim165\,$ks  of unfiltered exposure. 
The data were reduced using the software HEASOFT version $6.23$ and NICERDAS version 2018-03-01\_V004\footnote{https://heasarc.gsfc.nasa.gov/docs/nicer/nicer\_analysis.html}.
Because of the low intrinsic source flux and the relatively high spectroscopic sensitivity needed for the purposes of the present work, accurate background removal is important.
Good Time Intervals (GTIs) were first created applying standard filtering (e.g., removing events detected during South Atlantic Anomaly passages).
Then, further filtering was applied to remove high particle-radiation intervals associated with the Earth's auroral zones, i.e., ``the polar horns'', by applying a cut on the cutoff rigidity with {\tt COR\_SAX>4.0}.
In addition, detectors flagged as ``hot'' by the data analysis software were removed for each observation.
Then, GTIs were separated into times when \textit{NICER} was exposed to direct sunlight (orbit day), and times when
the satellite was within the Earth shadow (orbit night).
This procedure deals with different background components separately, such as the optical loading prominent only during day orbits at energies below $\sim0.35\,$keV.
In this way low energy data are free from artificial structures, although at the expense of a considerable amount of exposure time.
This does not represent the standard procedure for \textit{NICER} data analysis but ensures a low background level in the energy range $\sim0.3-0.5\,$keV, where an absorption feature is expected (see Section~\ref{sec:intro}).
Finally, a flat count rate cut (at $\sim10\,$c/s) to the resulting $0.25-12\,$keV light curve was applied to remove possible remaining background flaring events.
The resulting final exposures are $58\,$ks and $16\,$ks for day and night orbits, respectively.
The source spectra were grouped using the {\tt GRPPHA} tool to have a minimum of $25$ counts per bin.
The most recent response files provided by the \textit{NICER} instrumental calibration team were used\footnote{Response Matrix File and Ancillary Response File version $1.02$.}.

Background spectra were created from data acquired from one of seven ``blank sky'' targets based on the \textit{Rossi X-Ray Timing Explorer} (\textit{RXTE}) background fields \citep{Jahoda06}.
Among the seven available fields, we selected the target with the smallest angular separation from J1605 (BKGD \#8, $\Delta\theta\sim45^{\circ}$).
All observations of the background field were reduced as described above.
The resulting final exposures are $10\,$ks and $15\,$ks for day and night background spectra, respectively.

Source counts are $1.5\times10^5$ and $2.2\times10^5$, corresponding to a mean count rate of 5.9 and 3.6 s$^{-1}$ for night and day spectra, respectively, and accounting for about $94\%$ of the total.

\begin{deluxetable}{cccccc}
\setlength\tabcolsep{6.pt}
\tabletypesize{\scriptsize}	
\tablecolumns{8} 
\tablewidth{0pt}
\tablecaption{Observations log of RX~J1605.3+3249 \label{table:log}} 
\tablehead{
\colhead{Telescope} &\colhead{Obs ID} & \colhead{Start Time} & \colhead{Exposure\tablenotemark{a}} \\ 
      &  & [UTC] & [s] }
\startdata
\textit{NICER} & 1032020101 & 2017-07-19T22:46:40 &  115 \\
\textit{NICER} & 1032020102 & 2017-07-20T04:57:20 &  2620 \\
\textit{NICER} & 1032020103 & 2017-07-21T01:01:14 &  4545 \\
\textit{NICER} & 1032020104 & 2017-07-22T00:12:38 &  6038 \\
\textit{NICER} & 1032020105 & 2017-07-23T00:53:19 &  5924 \\
\textit{NICER} & 1032020106 & 2017-07-24T00:04:00 &  3184 \\
\textit{NICER} & 1032020107 & 2017-10-25T20:00:40 &  1111 \\
\textit{NICER} & 1032020108 & 2017-10-26T00:43:15 &  6324 \\
\textit{NICER} & 1032020109 & 2017-10-26T23:52:55 &  7141 \\
\textit{NICER} & 1032020110 & 2017-10-28T00:30:41 &  9187 \\
\textit{NICER} & 1032020111 & 2017-10-29T01:12:41 &  8565 \\
\textit{NICER} & 1032020112 & 2017-10-30T00:21:42 &  10246 \\
\textit{NICER} & 1032020113 & 2017-12-01T02:03:10 &  2529 \\
\textit{NICER} & 1032020114 & 2017-12-02T02:50:35 &  2367 \\
\textit{NICER} & 1032020115 & 2017-12-03T00:27:35 &  10105 \\
\textit{NICER} & 1032020116 & 2017-12-04T01:09:16 &  3535 \\
\textit{NICER} & 1032020117 & 2017-12-05T00:10:11 &  1587 \\
\textit{NICER} & 1032020118 & 2017-12-06T02:29:17 &  6060 \\
\textit{NICER} & 1032020119 & 2017-12-07T01:43:20 &  9660 \\
\textit{NICER} & 1032020120 & 2017-12-08T00:50:43 &  16545 \\
\textit{NICER} & 1032020121 & 2017-12-08T23:58:00 &  14728 \\
\textit{NICER} & 1032020122 & 2017-12-10T00:41:16 &  10856 \\
\textit{NICER} & 1032020123 & 2017-12-10T23:55:30 &  4672 \\
\textit{NICER} & 1032020124 & 2017-12-19T00:44:31 &  2681 \\
\textit{NICER} & 1032020125 & 2017-12-19T23:47:40 &  7175 \\
\textit{NICER} & 1032020126 & 2017-12-21T02:02:20 &  2700 \\
\textit{NICER} & 1032020127 & 2017-12-23T00:32:19 &  2944 \\
\textit{NICER} & 1032020128 & 2017-12-24T01:01:53 &  822 \\
\textit{NICER} & 1032020129 & 2017-12-26T19:56:44 &  143 \\
\textit{NICER} & 1032020130 & 2017-12-27T23:17:00 &  574 \\
\textit{NICER} & 1032020131 & 2018-04-01T05:57:20 &  186 \\
\textit{NICER}& 1032020132 & 2018-04-08T03:25:20 &  308 \\
\textit{XMM} &  0764460201 &  2015-07-21T20:19:26 &    121353  \\
\textit{XMM} &  0764460301 &  2015-08-20T18:07:20 &    68000  \\
\textit{XMM} &  0764460401 &  2015-08-20T18:07:20 &    73000  \\
\textit{XMM} &  0764460501 &  2016-02-10T22:36:56 &    62900  \\
\enddata
\tablenotetext{a}{Unfiltered times.}
\end{deluxetable}

\subsection{XMM-Newton}\label{subsec:xmm-data}

\textit{XMM} observed J1605 a total of four times between 2015 July 21 (Obs ID: 0764460201) and 2016 February 10 (Obs ID: 0764460501).
In this work, we only use data from the \textit{XMM} European Photon Imaging Camera (EPIC) cameras---i.e., the PN CCD \citep{Strueder01} and the MOS CCDs \citep{Turner01}---both sensitive in the $0.15-12\,$keV range and offering spectral resolution $E/\Delta\,E\sim20-50$ at $6.5\,$keV (see \citealt{Pires19}, for analysis of RGS data).
However, as J1605 has a soft spectrum, to avoid background contamination we restricted our analysis to the range $0.2-1.2\,$keV.
The PN and MOS cameras were set in Full and Large Window modes with thin filters, respectively.
The total unfiltered exposure was $\sim325\,$ks.
Data reduction was performed using the Science Analysis System (SAS) software \texttt{xmmsas\_20170719\_1539-16.1.0}, with the latest available calibration files.
Step-by-step reduction was performed following the official SAS Science Threads\footnote{https://www.cosmos.esa.int/web/xmm-newton/sas-threads}.
To remove high background flaring activity for EPIC cameras, single event (\texttt{PATTERN==0}), high energy light curves were extracted for each camera; 
then, a count-rate threshold was chosen corresponding to the low and steady background.
Applying such a threshold to the light curves resulted in the selection of Good Time Intervals.
For the pn-camera, single and double events were selected (\texttt{pattern${\leq}$4}), while single, double, triple, and quadruple events were accepted for the MOS cameras (\texttt{pattern${\leq}$12}).
Background circular regions of size $60^{''}$ to $100^{''}$ were defined off-source, on the same chip as the target for the PN camera, while on a different yet close and largely source-free chip for MOS, and used to generate background spectra.
PN and MOS spectra were extracted separately for each observation.

Because of the relatively high source flux, the thin filter, and the Full/Large Window observing mode, the observations are affected by pile-up at a few percent level in both PN and MOS cameras.
Even though the observed count rate is within ``tolerant'' levels according to \citet{Jethwa15}, we opted for a conservative approach in order to ensure a confident energy redistribution of the recorded events, thus helping in constraining spectral parameters and spectral features.
To minimize the resulting spectral distortion, we excised the core of the Point Spread Function (PSF) in each observation and extracted counts in an annulus centered on the source, retaining only the lower count-rate wings of the PSF.
To optimize the excising radius, we excluded progressively larger radii of the PSF core (up to a radius of about $15^{\prime\prime}$), testing their impact on the pile-up reduction using the \texttt{epatplot} task\footnote{https://www.cosmos.esa.int/web/xmm-newton/sas-thread-epatplot.} until pile-up effects were negligible.
This resulted in a loss of about 10 and 20\% of the original MOS and PN camera exposures, respectively.


We then used the SAS task \texttt{epicspeccombine} to combine spectra in order to improve statistics.
However, we notice that according to the SAS team the task can be used only to merge spectra and response files that have been generated in the same PI channel interval, but also that  selecting a spectrum with a non-standard PI range results in wrong response matrices and therefore unreliable spectra\footnote{\url{https://www.cosmos.esa.int/web/xmm-newton/sas-thread-epic-merging}}.
For this reason, we merged all the different cameras' spectra separately, ending up with one merged PN spectrum and one merged MOS spectrum.
The resulting final exposures are $285$ and $280$~ks for MOS2 and MOS1 and $225$~ks for PN. 
Spectral bins have been grouped to have a minimum of $25$ counts per spectral bin while the maximum oversample of the instrumental energy resolution was fixed to a factor of 3.

\section{Analysis}

\subsection{Pulsation Searches}\label{subsec:timing}
To search for a periodic pulsed signal associated with the rotation of the NS, the event detection times were first translated to the Solar System barycenter with the \texttt{barycorr} tool in HEASOFT for \textit{NICER} and the \texttt{barycen} task in SAS for \textit{XMM}. For this purpose we  adopted the DE405 Solar System ephemeris and the position of J1605 derived from the \textit{XMM} EPIC pn imaging data from the longest exposure (Obs ID 0764460201), RA=16:05:18.48, Dec=+32:49:21.0.

The periodicity searches were conducted using the PRESTO Fourier-domain pulsar search software package \citep{Ransom02}. The acceleration search technique implemented in PRESTO allowed us to coherently search long time series (up to several months) by considering a wide range of frequency drifts caused by a range of possible rotational spin-down values of the neutron star. 
For \textit{XMM}, only the EPIC PN data were used for this analysis, due to the significantly greater sensitivity and better time resolution ($\Delta t =73.4$ ms) of this instrument compared to the MOS cameras ($\Delta t= 0.9$ s). In addition, just a subset of the \textit{NICER} observations were used for the pulsation search---in particular, the observations taken in 2017 July 19--24 (22.4 ks of unfiltered exposure), 2017 October 25--30 (42.4 ks unfiltered exposure), and 2017 December 1--27 (82.7 ks unfiltered exposure), as they provide the most compact set of deep exposures, 
which is desirable for sensitive coherent pulsation searches. These \textit{NICER} subsets from July, October, and December data was first searched separately and also combined to perform a coherent search.

The \textit{XMM} events were binned at the intrinsic EPIC pn 73.4 ms detector time resolution, while the events from \textit{NICER} (which has an absolute time resolution of $\sim$100 ns) were binned at a time resolution of 0.977 ms ($1024$ Hz) for the separate searches of the 2017 July, October, and December subsets. To coherently search the combined October and December \textit{NICER} event lists, we used a 0.0625 s binning, and to search the July--December data set, we used a time binning of 0.25 s.

We first conducted periodicity searches over the 0.3--1.2 keV band. However, since J1605 may exhibit a multi-temperature thermal spectrum as suggested in previous works (e.g., \citealt{Pires+14}) and shown in Section~\ref{subsec:spectral_analysis}, it is possible that pulsations may only arise from the hotter and smaller regions on the stellar surface. Alternatively, the pulsations of the cool and hot emission regions may be significantly out of phase such that pulsations are strongly suppressed when integrated over the full 0.3--1.2 keV band \citep[see, e.g.,][for the curious case of the Puppis A pulsar]{Gotthelf09}. To account for this scenario, we conducted additional searches for periodicity  restricted to events in the 0.3--0.5 keV and 0.8--1.2 keV bands, where the cool and hot components dominate, respectively. 

The XINS RX J0720.4$-$3125 \citep{Borghese15} and RX J1308.6$+$2127 \citep{Borghese17} are known to exhibit narrow absorption features that only appear over a fraction of their rotation periods. In principle, the same could be the case for J1605 such that pronounced pulsations at the neutron star rotation period only occur in the absorption line. Motivated by this prospect, we conducted searches using only events in the energy range around the absorption feature apparent in the X-ray spectrum, 0.4--0.5\,keV.

No statistically significant ($\ge4\sigma$, as determined by the $Z^2_n$ test; see \citealt{Buccheri83}) periodic signals are found in either the \textit{NICER} or \textit{XMM} data sets for any choice of energy band. From the \textit{XMM} data we can set a 3$\sigma$ upper limit of 1.3\% on the pulsed fraction over the 0.3--1.2 keV band, assuming a sinusoidal pulse, for spin periods greater than $0.1468$\,s,  comparable to what \citet{Pires19} find.  For the \textit{NICER} data in the 0.3--1.2 keV band, after accounting for the additional number of trials from the acceleration search, we obtain a pulsed fraction limit of $<$2.6\% for spin periods greater than $1.95$\,ms from the 2017 December observations, a $<$1.6\% limit for periods greater than $0.125$\,s by combining the 2017 October and December observations. 
Including the \textit{NICER} exposures from 2017 July 19--24 data does not result in a significant improvement in sensitivity to pulsations due to the 3 month gap and shorter exposure relative to the October and December data, yielding a $<$1.6\% limit for periods $>$0.5 s. The $<$2.6\% limit (at $3\sigma$) for short periods ($>$2~ms) is significantly stronger than the 3.2--5.0\% limits (at $2\sigma)$ for periods $>$1.2~ms obtained by \citet{Kerkwijk04} using \textit{XMM} pn fast timing data (see in particular their Table 3).

The absence of pulsations in substantially deeper exposures indicates that the period reported in \citet{Pires+14} based on a 60 ks \textit{XMM} exposure was spurious, in agreement with \citet{Pires19}.


\subsection{Spectral analysis}\label{subsec:spectral_analysis}

\begin{table}[!t]
\caption{Best-fit results of J1605 spectral analysis with a double-blackbody model and two atmospheric models, \texttt{NSA} and \texttt{NSMAXG}. All reported errors are at $90\%\,$c.l.} \label{table:spectral}
\begin{ruledtabular}\begin{tabular}{lccc}
 & Double-BB & NSA $(g=2.43$) & NSMAXG\\
N$_{\textrm{H}}$ [$10^{20}\,$cm$^{-2}$] & $4.6^{+1.4}_{-1.4}$ &  $3.4^{+0.2}_{-0.2}$ & $5.6^{+0.3}_{-0.3}$\\
kT$_\textrm{cool}\,$[eV] & $63^{+7}_{-6}$ & --& --\\
kT$_\textrm{hot}\,$[eV] & $119^{+6}_{-4}$ & --& --\\
log\,T$_\textrm{eff}\,$[K] & -- & $5.737^{+0.005}_{-0.005}$ & $5.729^{+0.016}_{-0.022}$\\
M$_\textrm{NS}\,$[M$_{\odot}$] & -- & 1.4  (fixed) & $2.04^{+0.19}_{-0.49}$ \\
R$_\textrm{NS}\,$[km] & -- & 10 km (fixed) & $15.6^{+0.62}_{-0.79}$\\
B\,[$10^{13}\,$G] & -- & 1 (fixed) & 1 (fixed) \\
K$_\textrm{cold}$ & $3.3^{+8.0}_{-2.2}\,\textrm{E+}5$ & -- & --\\
K$_\textrm{hot}$ & $1.9^{+0.8}_{-0.8}\,\textrm{E+}3$ &  --& --\\
Distance$^a$ [kpc] & $0.174^{+0.127}_{-0.079}$ &  $0.092^{+0.005}_{-0.005}$ & 0.1 (fixed)\\
K$_\textrm{atmos}$ & -- & $1.19^{+0.12}_{-0.12}\textrm{E-}4$ & 1.0 (fixed)\\
E$_\textrm{abs}\,$[keV] & $0.435^{+0.013}_{-0.006}$ & $0.452^{+0.003}_{-0.003}$ & $0.445^{+0.003}_{-0.003}$\\
$\sigma_\textrm{abs}\,$[keV] & $0.110^{+0.010}_{-0.011}$ & $0.087^{+0.004}_{-0.005}$ & $0.092^{+0.005}_{-0.004}$\\
$\tau_\textrm{abs}$ & $0.24^{+0.13}_{-0.08}$ & $0.098^{+0.009}_{-0.008}$& $0.133^{+0.011}_{-0.016}$\\
C$_{MOS}$$^b$ & $0.849^{+0.006}_{-0.005}$ &$0.849^{+0.006}_{-0.005}$ & $0.857^{+0.006}_{-0.006}$  \\
C$_{Night}$$^c$ & $0.848^{+0.013}_{-0.013}$ & $0.848^{+0.013}_{-0.013}$ & $0.858^{+0.013}_{-0.013}$\\
C$_{Day}$$^d$ & $0.725^{+0.005}_{-0.005}$ & $0.725^{+0.005}_{-0.005}$& $0.733^{+0.005}_{-0.005}$\\
Flux$^e$& $1.10^{+0.42}_{-0.27}$ & $1.08^{+0.04}_{-0.03}$ & $1.16^{+0.40}_{-0.34}$\\
$\chi^2_\textrm{red}$/d.o.f. & $1.12/483$ & $1.15/485$& 1.14/484\\
\end{tabular}\end{ruledtabular}
\tablenotetext{}{Parameters indicated with $K_{x}$ represent the normalization value of the \textit{cold} and \textit{hot} component for the double-blackbody model, and that of the \textit{atmospheric} \texttt{NSA} and \texttt{NSMAXG} models. \qquad $^a$ Distance D of the source calculated as proportional, or equal to, $1/K_x^2$ for the double-BB and \texttt{NSA} model, respectively (see text), or reported as a parameter of the model for the \texttt{NSMAXG} model.
\quad $^{b,c,d}$ Cross-normalization factors for \texttt{MOS}, \textit{NICER} night and day spectra, respectively. \texttt{PN} cross-normalization factor was kept fixed to unity.
\quad $^e$ Unabsorbed flux calculated for the continuum component(s) in the $0.3-1.2\,$keV band and reported in units of $10^{-11}\,$erg\,cm$^{-2}\,$s$^{-1}$. Flux values with estimated errors were derived using the \texttt{cflux} model from \texttt{XSPEC}.}
\end{table}

Spectra were fitted using the \texttt{XSPEC 12.10.0} package \citep{Arnaud96}.
We fitted \textit{NICER} (day and night) and \textit{XMM} (pn and MOS) spectra simultaneously, allowing for a cross-normalization factor among the different spectra.
The energy band for fitting was limited to $0.2-1.2\,$keV, above which the background dominates.

Photoelectric absorption model and elemental abundances were set according to \citet{Wilms00} (\texttt{tbabs} in \texttt{XSPEC}) to account for photoelectric absorption by neutral interstellar matter (or column density N$_\textrm{H}$), and assuming model-relative (\textit{wilm}) solar abundances.
During our spectral analysis we allowed the column density parameter to vary (although we also explored the case where the absorption column density was kept fixed to the Galactic value, $2.4\times10^{20}\,$cm$^{-2}$).
This resulted in values larger than the Galactic N$_\textrm{H}$ value to the source ($\sim3-5\times10^{20}\,$cm$^{-2}$), as well as larger than the value found in previous works (e.g., \citealt{Kerkwijk07,Pires+14}).
This is similar to what \citet{Pires19} also found, i.e. nominal values of the N$_\textrm{H}$ parameter generally higher than Galactic (that is, in the range $2.5-5.3\times10^{20}\,$cm$^{-2}$).
Moreover, as discussed in \citet{Vigano14}, the single-blackbody model used in previous works to fit the X-ray data of J1605 (see Section~\ref{sec:intro}) tends to underestimate N$_\textrm{H}$  by $20-30\%$ with respect to the actual value. 
However, the higher than Galactic N$_\textrm{H}$ value in the case of J1605 is related to the co-variance with the parameters of the broad absorption line at $\sim0.4\,$keV and to our ignorance of the true continuum spectral shape.
We also verified that employing different absorption models (e.g., \texttt{wabs} in \texttt{XSPEC}, see \citealt{Morrison83}) and abundances (e.g., \texttt{angr} in \texttt{XSPEC}, see \citealt{Anders89}) returns consistent N$_\textrm{H}$ values to those obtained using the \texttt{tbabs} model.

The continuum of J1605 is known to be fitted by an absorbed single- or double-blackbody model modified by at least one Gaussian absorption line at $0.45\,$keV \citep{Kerkwijk04,Pires+14} and we tested both models to fit our data.
We used the \texttt{bbodyrad} model (instead of the simpler \texttt{bbody}) from \texttt{XSPEC} because it allows to link the normalization $K$ of the blackbody component to the emitting surface area through the relation $K=R_\textrm{km}^2/D_\textrm{10\,kpc}^2$, where $R_\textrm{km}$ is the radius of the emitting surface in units of km and $D_\textrm{10\,kpc}$ is the distance from the source in units of $10\,$kpc.
A single-blackbody model returns a reduced $\chi^2>5$ (for $488$ d.o.f.).
Our data are best fit by an absorbed double-blackbody model with an absorption feature at $\sim0.43\,$keV (see Table~\ref{table:spectral}).
The absorption feature has been modeled with a Gaussian profile (the multiplicative \textit{gabs} component in XSPEC):

\begin{equation}\label{gabs}
F(E)=\text{exp}\left\{-\frac{\tau_{abs}}{\sqrt{2\pi}\sigma_{abs}}\text{exp}\left(-\frac{(E-E_{abs})^2}{2\sigma_{abs}^2}\right)\right\},
\end{equation}

where $E_{abs}$ is the line centroid energy, $\tau_{abs}$ and $\sigma_{abs}$ are the optical depth and the width of the line, respectively.
The second blackbody component and the absorption feature improve the reduced $\chi^2$ to a value of $2.4$ and $1.1$, respectively. 
The blackbody temperatures and absorption line energy are in general agreement with findings from \citet{Pires19}.
On the contrary, previous works (e.g., \citealt{Haberl07,Pires+14}) require additional absorption features at $\sim0.58$ and $0.8\,$keV.
However, we note that the observation of spectral features in the spectra of XINS can be model-dependent, and can also depend on the exact details of the analysis.
Moreover, previous results based on \textit{XMM} observations of J1605 are now superseded by \citet{Pires19} analysis and data set.
On the other hand, we notice that results from \citet{Haberl07} are obtained using a single-temperature blackbody model and can not therefore be compared to that used in the present work.
Moreover, the fit obtained by \citet{Haberl07} is only marginally acceptable, with a reduced $\chi^2_\textrm{red}\sim1.4$, and structured residuals resulting from the best-fit model.

The absorption feature found in previous works at $\sim0.58\,$keV \citep{Kerkwijk04, Hohle12,Pires+14} is narrow (Gaussian width in the range $\sim3-16\,$eV) and needs the resolution power of instruments like RGS on board \textit{XMM} to be resolved, which is beyond the scope of the present work.
However, we note that \citet{Pires19} also found no evidence of the narrow absorption feature at $\sim0.58$ keV in the most recent \textit{XMM}-RGS observations.

Finally, we note that \citet{Schwope07} find that the energy of the two absorption features observed in the XINS RBS~1223 spectrum considerably changes among different observations, with the lowest energy absorption feature changing from $0.39\,$keV in $2003$ to $0.20\,$keV in $2005$ and the higher energy absorption feature going from undetected to $0.73\,$keV, respectively (while roughly harmonic in the rest of the observations). 
Therefore, the centroid energy of the detected features may vary on time scales of years (a somewhat analogous phenomenon is observed in accreting NSs, e.g. Her X-1 \citealt{Staubert16}, although the accretion process in those sources is likely responsible for the long-term variation of the cyclotron line energy).
More recent works \citep{Hambaryan11,Borghese17} also find inconsistent line energies for RBS 1223, although the inconsistency might be due to the difference in the analysis approach.

More complex, physical models have also been tested, such as those representing the X-ray spectrum emitted from the atmosphere of a NS.
Various NS atmospheric models have been proposed in the literature, and tested in the present work, e.g., the magnetic/non-magnetic versions of the fully ionized hydrogen atmosphere model (\texttt{NSA} in \texttt{XSPEC}, \citealt{Zavlin96, Pavlov95}), the non-magnetic hydrogen atmosphere model with variable surface gravitational acceleration (\texttt{NSAGRAV} in \texttt{XSPEC}, \citealt{Zavlin96}), and the weakly/strongly magnetized versions of the partially ionized atmospheric model that allows for variable surface gravitational acceleration and is composed of hydrogen (H) or heavier elements (e.g., carbon, oxygen, iron, \texttt{NSX} and \texttt{NSMAXG} model in \texttt{XSPEC}, \citealt{2009Natur.462...71H,Ho14}).
Among all tested atmospheric models, the only ones that returned an acceptable fit were (1) the \texttt{NSA} model in the magnetic case (B${=}10^{13}\,$G); and (2) the \texttt{NSMAXG} model with B${=}10^{13}\,$G, the latter with the distance and normalization values kept fixed (see Table~\ref{table:spectral}).
None of the other tested combinations of the above mentioned atmospheric models returned a statistically acceptable fit or physically meaningful values of the model parameters, and will not be discussed further.
Moreover, we stress that the source distance commonly adopted in the literature ($\sim350\,$pc; see, e.g., \citealt{Posselt07}) is nonetheless uncertain and cannot be readily accommodated by our data, while a possible distance of $\sim150\,$pc (see Section~\ref{sec:intro}) is consistent with our data.

Similarly to the double-blackbody model, the atmospheric model fits also require a Gaussian absorption line at $\sim0.45\,$keV.
Finally, because the \texttt{NSA} model is developed for a standard gravitational acceleration $g=2.43\times10^{14}\,$cm\,s$^{-2}$, corresponding to standard NS mass M$_{\textrm{NS}}=1.4\,$M$_{\odot}$ and radius R$_{\textrm{NS}}=10\,$km, these two parameters were kept fixed during the fitting procedure\footnote{https://heasarc.gsfc.nasa.gov/xanadu/xspec/manual/node196.html}.

Analogous spectral continuum models and spectral features have been employed by \citet{Pires19} to fit \textit{XMM} data of J1605.
Their spectral results are overall consistent with those found in the present work.
However, their best-fit spectral models of EPIC data find an absorption feature at $385\,$eV, a significantly different energy than that found here, while the energy of the absorption feature found by the analysis of \textit{XMM}-RGS spectra is consistent with our results.

\section{Discussion}

\subsection{A double-blackbody model for the continuum emission}\label{subsec:bbody}

\begin{figure}[!t]
\includegraphics[width=0.45\textwidth]{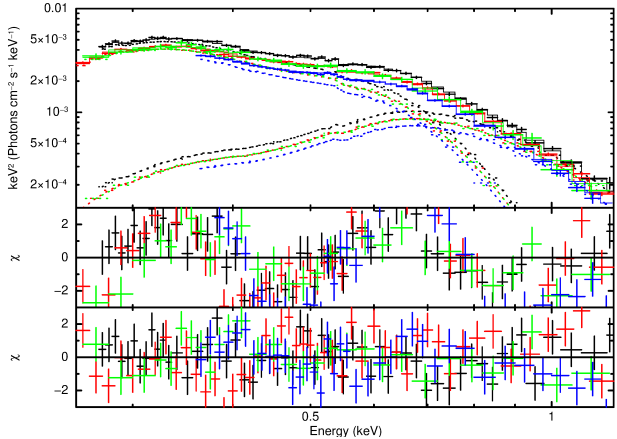}
\caption{\textit{NICER} and \textit{XMM} spectra of J1605 fitted with an absorbed double-blackbody continuum plus a Gaussian absorption line.
\textit{Top panel:} Data and folded model of the \textit{XMM} (PN and MOS -- black and red, respectively) and \textit{NICER} (night and day -- green and blue, respectively) spectra of J1605.
Dotted lines represent the cold and hot components (peaking at softer and higher energy, respectively).
\textit{Central panel:} residuals for the double-blackbody model without the Gaussian absorption component.
\textit{Bottom panel:} residuals for the best-fit model. Spectra and residuals have been rebinned for plotting purpose.
}
\label{fig:bbodies_spec}
\end{figure}

The spectral continuum of XINS is broadly consistent with a single-temperature blackbody component ($kT_{bb}\sim40-100\,$eV) slightly modified by interstellar absorption (N$_\textrm{H}\sim10^{20}\,$cm$^{-2}$).
This result is generally interpreted as thermal emission from the NS surface.
However, with the availability of high signal-to-noise spectra, deviations from that simple model emerge (see, e.g., \citealt{Zane11} and references therein).
Phase-averaged as well as phase-resolved spectra from XINS generally are better fitted by a combination of two or three blackbody components, physically interpreted as coming from different regions of the star surface.
Such deviations from a purely single-temperature component are expected as the result of an inhomogeneous temperature distribution across the NS surface, expected from theoretical arguments and stemming from the presence of strong magnetic fields causing, e.g., anisotropic thermal conductivity and non-spherically symmetric magnetic field dissipation, leading to the presence of hot spots \citep{Page07,Pons09}.
In particular, the origin of the double-blackbody continuum is generally attributed to two thermally emitting spots: a smaller hot spot, usually associated with the magnetic poles, and a larger cool spot as wide as the NS itself and due to the cooling surface.

\begin{figure}[!t]
\centering
\includegraphics[width=0.45\textwidth]{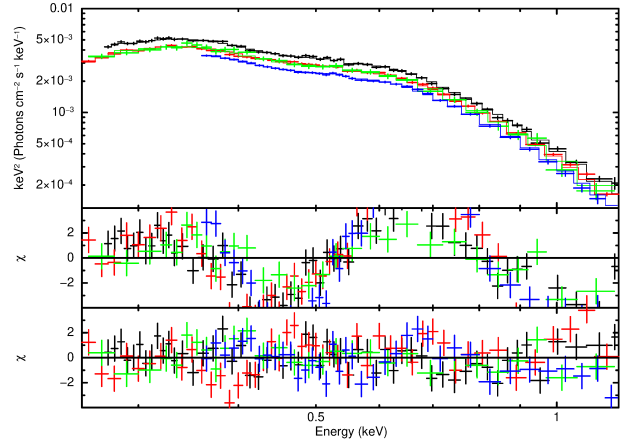}
\caption{\textit{NICER} and \textit{XMM} spectra of J1605 fitted with an absorbed magnetized hydrogen atmospheric (\texttt{NSA}) model plus a Gaussian absorption line.
\textit{Top panel:} Data and folded model of the \textit{XMM} (PN and MOS -- black and red, respectively) and \textit{NICER} (night and day -- green and blue, respectively) spectra of J1605.
\textit{Central panel:} residuals for the pure \texttt{NSA} model without the Gaussian absorption component.
\textit{Bottom panel:} residuals for the best-fit model.
Spectra and residuals have been rebinned for plotting purpose.}
\label{fig:NSA_spec}
\end{figure}

\begin{figure*}[t]
\centering
\includegraphics[width=\textwidth]{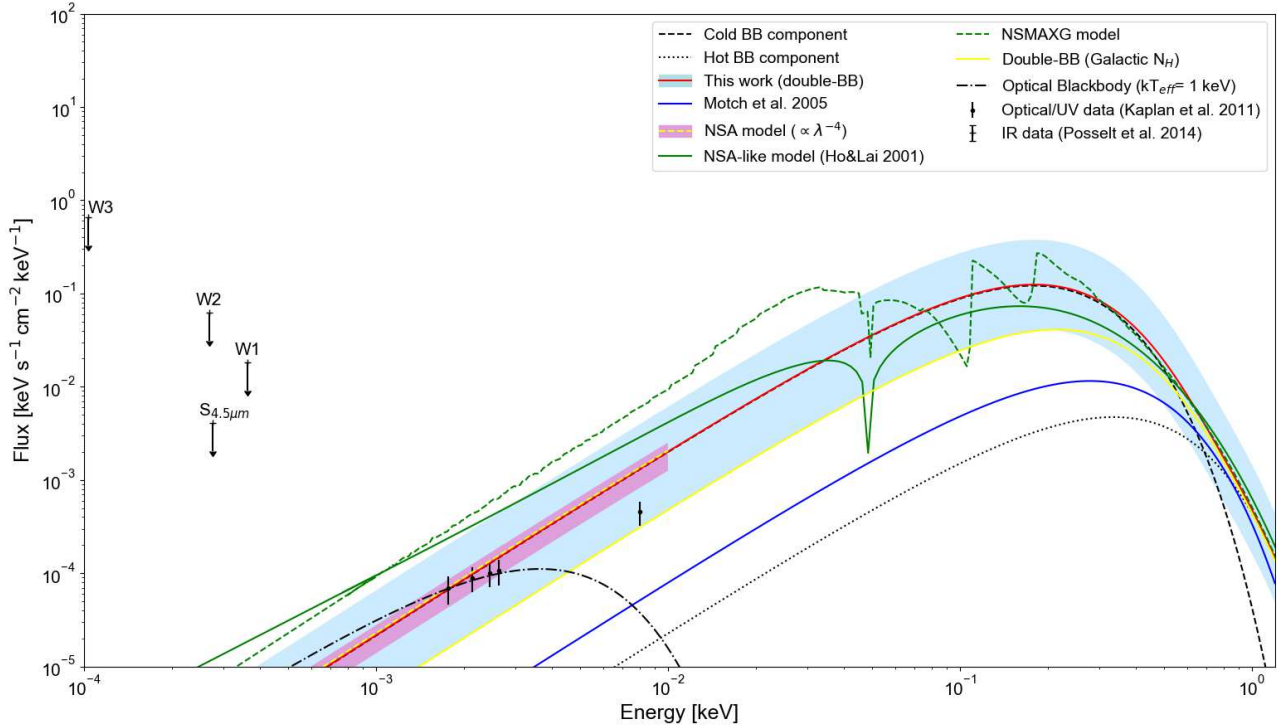}
\caption{Broadband spectral energy distribution of J1605.
Black dotted and dashed lines are the hot and cold (unabsorbed) blackbody components, respectively, obtained from the fit of X-ray data (see Table~\ref{table:spectral}).
The red continuous line represents the sum of the two blackbody components (uncertainty at $90\%$ c.l. shown as the cyan shaded region).
The yellow solid line represents the best-fit double-blackbody model obtained with N$_\texttt{H}$ fixed to the Galactic value to the source.
For comparison, the best-fit single-blackbody model obtained by \citet{Motch05} is also shown (continuous blue line).
IR upper limits are shown as black arrows. Optical/UV data are shown as black points with error bars, while the black dash-dotted line represents blackbody emission at kT$_{\rm eff}=1\,$keV (see text and \citealt{Ertan17}).
The best-fit \texttt{NSA}-like model \citep{Ho+Lai01} and \texttt{NSMAXG} model are also reported for comparison (continuous and dashed green lines, respectively).
The low-energy tail predicted by the \texttt{NSA} model corrected for a color factor of $2.5$ is shown (yellow dashed line), including propagated uncertainty (plum shaded region) in the relevant energy band.}
\label{fig:bbodies}
\end{figure*}

Our analysis finds that a double-blackbody model fits well the continuum emission from J1605 (see Figure~\ref{fig:bbodies_spec}).
The temperatures of the cold and hot regions are $63$ and $119\,$eV, respectively, in general agreement with previous work on J1605 and other XINS.
Assuming a $10\,$km-radius NS for the cold component (see Section~\ref{subsec:NSA}) we get a distance of $\sim174^{+127}_{-80}\,$pc, while the normalization of the hot component, given a nominal distance of $174\,$pc returns a radius of the emitting hot spot equal to $\sim0.76^{+0.18}_{-0.18}\,$km.
The case of the double-blackbody model with the column density value fixed to the Galactic value (N$_{\textrm{H}}=2.4\times10^{20}\,$cm$^{-2}$) has also been investigated.
This model fits well the X-ray data ($\chi^2_{red}/d.o.f. = 1.14/484$) and returns a distance of $393\pm18\,$pc, more in line with previous works \citep[and references therein, see also Sect.~\ref{sec:intro}]{Tetzlaff12}, while both the hot and cold blackbody components show about $10\,$eV hotter temperature values.

Finally, to investigate deviations from a pure, single-blackbody continuum as found in \citet{Motch05} for J1605, we compared that with the broadband spectral energy distribution obtained from our double-blackbody model.
Our results are presented in Fig.~\ref{fig:bbodies}.
Below $\sim0.5\,$keV, the flux predicted by the double-blackbody model is about $10$ times higher than that predicted by the single-blackbody, while the hot blackbody component returns a spectrum that is roughly comparable to that of the single-blackbody model of \citet{Motch05}, considering the relatively small difference between the temperature values ($119\,$ and $99\,$eV, respectively).
However, the cold component brings an important contribution to the soft X-ray band ($\lesssim0.5\,$keV), modifying the emerging spectrum at softer X-ray energies.

An important consequence deriving from the inclusion of a cold component is the model prediction at lower wavelengths, namely in the IR/optical/UV band.
Fig.~\ref{fig:bbodies} compares the prediction of our model with Hubble Space Telescope (\textit{HST})/\textit{Subaru} optical/UV data from \citet{Kaplan03,Motch05,Kaplan11} as well as \textit{WISE} and \textit{Spitzer} IR data \citep{Posselt14}.
Optical/UV magnitudes have been converted to flux according to the standard STDMAG conversion \citep{Kaplan03,Kaplan11}, and corrected to account for optical extinction following the empirical relation described in \citet{Foight16}, N$_{\textrm{H}} = (2.87\pm0.12)\times10^{21}$ A$_{\textrm{V}}\,$cm$^{-2}$ (which, however, suffers from considerable scatter at low N$_{\textrm{H}}$ values such as those derived for J1605).
The model can fit optical data from the B to the R band without requiring additional components or conditions, such as a power law component or a thin hydrogen atmosphere \citep{Motch03,Motch05,Ho+07}.
This is a direct result of the inclusion of a second, colder blackbody component in the fit.
However, as illustrated by \citet{Kaplan11}, J1605's optical/UV data show a trend that is less steep than a $\propto\lambda^{-4}$, blackbody-like function.
As a result, the nominal double-blackbody model fits the optical data but results in an overestimation of the UV flux, consistent with it only within the large uncertainty (see Figure~\ref{fig:bbodies}).

\subsection{The fallback disk scenario}

As outlined in Sect.~\ref{subsec:bbody} and Fig.~\ref{fig:bbodies}, the best-fit double-blackbody model is subject to further considerations.
If the physical interpretation of the double-blackbody model is plausible, and only the nominal predicted values are considered, then a mechanism must be at work to suppress the inferred flux in the UV band.
A possible mechanism responsible for the suppression of the UV flux in XINS is due to potential fallback disks or dusty belts surrounding the compact object which are, at least in some cases, indicated as a distinct possibility \citep{Perna00,Posselt14,Posselt18}. 
When present, dust grains around the NS are heated by the UV radiation, for which the grains behave as nearly perfect absorbers, thus reradiating the absorbed flux in the infrared band.
However, the double-blackbody model predicts a flux that is ${\sim}3$ times larger than that observed in UV (see Fig.~\ref{fig:bbodies}), which requires absorption values of 
a few times $10^{21}\,$cm$^{-2}$, that is about 10 times the Galactic value to the source, and many times the amount observed in our best-fit model (see Table~\ref{table:spectral}).
Assuming that the fallback disk is distributed as a torus with inner and outer radii of $10^{10}$ and $10^{16}$ cm, respectively (see \citealt{Posselt14}), a column density of $3\times10^{21}\,$cm$^{-2}$ corresponds to a torus mass of about $30\,$M$_\Earth$.
The IR emission from such a massive torus would therefore be intense, e.g. according to Eq.~1 in \citet{Posselt14}, of the order of $10^2\,$mJy at $160 \micron$ and for a source distance of $173\,$pc. 
However, \citet{Posselt14} found no significant infrared counterpart at the position of J1605, a result that puts upper limits on the IR emission from this source, $<12.2$ mJy at $160\, \micron$ (that is, approximately $18\,$keV\,s$^{-1}\,$cm$^{-2}\,$keV$^{-1}$ at $8\times10^{-6}\,$keV, see Fig.~\ref{fig:bbodies} for comparison) and on the possible disk mass, $<2.2\,$M$_\Earth$.
Therefore, a dusty fallback disk surrounding the NS seems unsuitable to explain the UV suppression implied by the double-blackbody model.
Alternatively, the best-fit double blackbody model with the column density value fixed to the Galactic value can be considered.
Fig.~\ref{fig:bbodies} shows that this model is consistent with the UV data point (at $\sim10^{-2}\,$keV), but not with the optical data. 
This result can be interpreted in terms of the model employed by \citet{Ertan17}, where the optical flux is emitted mainly from the inner rim of the fallback disk in the form of a blackbody spectrum.
In Fig.~\ref{fig:bbodies} we show a blackbody model with kT$_\textrm{eff}\sim1.3\,$keV and optical flux F$_\textrm{opt}\sim10^{-3}\,$F$_\textrm{X}$, where F$_\textrm{X}$ is the X-ray flux.
These values are typical of XINS \citep{Ertan17} and result to fit well the optical data leaving the UV flux almost unaffected.

\subsection{An atmospheric model for the continuum emission}\label{subsec:NSA}

Besides the double-blackbody model, X-ray spectra from J1605 analyzed in this work are equally well fit by specific configurations of the atmospheric \texttt{NSA} and \texttt{NSMAXG} models (see Table~\ref{table:spectral}).
In our analysis, both models assume constant magnetic field strength and temperature across the NS surface.
While one could expect magnetic field (and temperature) variation across the emission region, the non-detection of pulsations (see Section~\ref{subsec:timing}) supports our assumption.

The \texttt{NSA} model provides the effective unredshifted temperature of the NS surface and can be fitted for a few (fixed) values of the magnetic field, while providing the source distance as a free parameter.
Our data are well fitted by the \texttt{NSA} model with B$=10^{13}\,$G. 
Similar to the double-blackbody model employed in Section~\ref{subsec:bbody}, the normalization factor K$_{\rm atmos}$ of the \texttt{NSA} model is linked to the distance $d$ (in units of parsec) through the relation $K_{NSA}=1/d^2$, thus resulting in a distance of $92^{+5}_{-5}\,$pc, consistent with the results from Section~\ref{subsec:bbody} and those hinted by \citet{Motch99,Motch05}.
The surface temperature returned by this model is $\log\,$T$_{\rm eff}=5.737\pm0.005\,$K ($47.0\pm0.5$ eV).

A constrained configuration of the \texttt{NSMAXG} model has also been found to fit the data.
The best-fit \texttt{NSMAXG} model consists of a magnetized (B$=10^{13}\,$G) NS at a fixed distance of $100\,$pc and a normalization value, equal to the ratio of the emitting region compared to the NS radius $($R$_\textrm{em}/$R$_\textrm{NS})^2$ fixed to unity.
This model returns a similar surface temperature as that of the \texttt{NSA} model, $\log\,$T$_{\rm eff}=5.729^{+0.016}_{-0.022}\,$K ($46.2^{+1.7}_{-2.3}\,$eV), and a ``scaled-up'' version of the standard NS, with $M_{\textrm{NS}}=2.04\,$M$_{\odot}$ and R$_{\textrm{NS}}=15.6\,$km (corresponding to a gravitational redshift of $z_g=0.28$).

In analogy with Section~\ref{subsec:bbody}, the \texttt{NSA} and \texttt{NSMAXG} models also have been tested with the absorption column density fixed to the Galactic value.
However, these models do not fit the data well, leaving enhanced wave-like residuals ($\chi^2_\textrm{red} = 1.4$), especially in the softest ($<0.35\,$keV) part of the spectrum, and will not be further discussed.

Although atmospheric models have been found to satisfactorily fit the X-ray emission of XINS in previous work as well, the temperature values so obtained generally overestimate the observed optical flux by a factor of $\sim10-100$ (see, e.g., \citealt{Pavlov96,Pons09,Motch03,Burwitz03}).
For this, we compare the \texttt{NSA} model predictions with the optical/UV data provided in \citet{Motch05,Kaplan03,Kaplan11}, as well as with the double-blackbody model (see Section~\ref{subsec:bbody}).
In fact, at energies softer than the X-ray domain, the blackbody and the \texttt{NSA} models both follow a power-law, $F(\lambda)\propto\lambda^{-4}$ (see also \citealt{Ho08}).
However, the \texttt{NSA} model spectra implemented in \texttt{XSPEC} only extends down to $0.05\,$keV.
Furthermore, at $B=10^{13}$~G, the proton cyclotron spectral feature occurs at $0.063\,$keV and significantly distorts the continuum spectrum near this energy (see Figure~\ref{fig:bbodies}). 
This prevents extending the \texttt{NSA} model to optical wavelengths in the form of a power-law.
Thus, in order to illustrate a fully ionized hydrogen atmosphere spectrum at optical wavelengths, we compute and show an analogous spectrum (i.e., a \texttt{NSA}-like model) using the method described in \citet{Ho+Lai01} and values obtained from the \texttt{NSA} best-fit (see Table~\ref{table:spectral}).
As shown in Fig.~\ref{fig:bbodies}, the \texttt{NSA}-like model is consistent at optical wavelengths with the double-blackbody model within its relatively large uncertainty. This is still noticeable, since other XINS show high discrepancy between the two models (as reported above). 
On the other hand, the best-fit \texttt{NSMAXG} model shows a discrepancy from the nominal double-blackbody at optical wavelengths by a factor of almost 5.
Note that the \texttt{NSMAXG} spectrum shown in Figure~\ref{fig:bbodies} takes into account the dense-plasma effect described by \citet{2003ApJ...599.1293H}, which occurs when photons of frequency below the local plasma frequency have their propagation hindered; this effect causes the spectrum to deviate from a $\lambda^{-4}$ behavior but the precise nature of the deviation is uncertain, and therefore the spectrum shown here is not definitive.

Finally, in Figure~\ref{fig:bbodies} we also show the low-energy tail predicted by the \texttt{NSA} model ($\propto\lambda^{-4}$) approximated as a blackbody spectrum (dashed yellow line) whose temperature is equal to that found by the best-fit model of the \texttt{NSA} component ($\log\,T_\textrm{\rm eff}=5.737\,$K, see Table~\ref{table:spectral}) and corrected by a color factor of $2.5$\footnote{
This value of the color factor is obtained by \citet{Zavlin96} for an atmospheric spectrum with the same temperature and composition (pure hydrogen) of the NSA model considered here (see Figure 5 of their work).}.
Despite the fact that color factors found by those authors and employed here are computed for non-magnetic models, it is remarkable how well the resulting spectrum meets the double-blackbody model at optical/UV wavelengths.
The \texttt{NSA} model so obtained, similarly to the nominal double-blackbody model, is consistent with the optical data but not with the UV data, leading to a possible similar interpretation (see Section~\ref{subsec:bbody}).

\subsection{A pCF origin for the absorption feature}\label{subsec:pcrsf}

Spectral absorption features are commonly observed among XINS.
These features are usually attributed to electron or proton cyclotron interactions, and/or to electronic transitions in partially ionized or condensed atmospheres (see Section~\ref{subsec:atmos_abs}).
However, absorption features have been also shown to result spuriously as a consequence of fitting, e.g., multi-temperature blackbody emission models \citep{Vigano14}.

A cyclotron resonant feature is expected in NSs with high ($B\geq10^{12}\,$G) magnetic fields, where the electron/proton motion perpendicular to the magnetic field lines is quantized in discrete Landau levels, and so are the energies corresponding to those levels, thus resulting in resonant scattering of photons at those energies.
The energy of the fundamental line is 

\begin{equation}
E_\textrm{CRSF}\approx\frac{11.6}{(1+z_g)}\frac{m_e}{m_x}\,B_{12}\,\textrm{\,keV}
\end{equation}
where $B_{12}$ is the magnetic field in units of $10^{12}\,$G, $m_e$ and $m_x$ are the mass of the electron and that of the particle responsible for the photon scattering, respectively, and $z_g$ the gravitational redshift ($\sim0.3$ for standard NS mass and radius).

In our analysis, the absorption feature in the spectrum of J1605 is found at a nominal energy of $0.432-0.451\,$keV (depending on the best-fit model, see Table~\ref{table:spectral}).
Assuming canonical values for the NS mass and radius ($M=1.4\,M_\odot$ and $R=10\,$km) and an average value of the absorption line of $0.44\,$keV, the resulting magnetic field is $9.0\times10^{13}$ and $4.9\times10^{10}\,$G for the proton and electron features, respectively.

Timing studies point out that all other pulsating XINS harbor a magnetic field of the order of $10^{13}\,$G, an order of magnitude that is consistent with the results from our fit of the atmospheric model and that inferred by the pCF (see Table~\ref{table:spectral} and Section~\ref{subsec:NSA}).
Moreover, we notice that in the case of pCF, the feature is expected to be narrower, with line widths of the order of hundreds of eV (\citealt{Nishimura03}, and references therein), contrary to widths of the order of keV for eCRSF, although line strength suppression by vacuum polarization in high magnetic fields plays a role in both features (see, e.g., \citealt{Ho+Lai03}).
We observe a line width of $\sim110\,$eV, in agreement with expectations for a pCF and comparable to the width of Gaussian absorption lines in other XINS whose independent measurements of the magnetic field favor the pCF interpretation (see, e.g., \citealt{Cropper04}).
Therefore, if the origin of the absorption feature is to be ascribed to cyclotron resonance, our analysis tends to favor the proton rather than the electron as the particle responsible for the scattering/absorption process.

Previous work claimed other absorption features in the spectrum of J1605 that have not been detected in the present work \citep{Haberl07,Pires+14}.
These features had centroid energies with a $2{:}3{:}4$ ratio, and were therefore interpreted as the result of harmonic cyclotron features.
In Section~\ref{subsec:spectral_analysis} we gave technical reasons that might explain the different results presented in this work.
Here we point out a further physical reason that highlights the difficulty to observe proton harmonic cyclotron resonant features in the spectra of highly-magnetic, thermally-emitting INSs.
In fact, in such physical conditions the strength of each pCF harmonic would scale with the feature centroid energy $E$ as $\sim E/m_xc^2$ with respect to the fundamental (where $m_x$ is the mass of the particle responsible for the scattering).
Accordingly, pCFs would result in progressively weaker lines \citep{Kerkwijk07,Schwope07,Potekhin10}, contrary to the observations.
However, we note that quantum effects can make electron cyclotron harmonics stronger than the simple $\sim E/m_xc^2$ scaling \citep{Suleimanov10,Suleimanov12}, and therefore make them possible to be observed.
Furthermore, \citet{Pires19} also found no evidence of the previously reported absorption features.

Finally, as noted by \citet{Vigano14}, surface temperature inhomogeneities can mimic absorption lines, at least in the spectra of some XINS.
Those authors find that a pure blackbody model can result in a spurious absorption feature at $\sim0.45\,$keV which, instead, can be accounted for by a synthetic model composed of several surface temperature distributions.
However, the employment of a double-blackbody model should ensure our analysis to be free from spurious detections at $\sim0.45\,$keV because this energy value lies close to the Wien peak of one of the blackbody components, thus enhancing the robustness of our results.
Nonetheless, a double-blackbody model also can result in spurious absorption features around the energy where the flux from the cold and hot components is comparable.
For J1605, our analysis reveals that comparable flux contributions in the double-blackbody model are found between $0.7-0.8\,$keV (see Figure~\ref{fig:bbodies_spec}), consistent with the centroid energy of the broad absorption line reported in previous works on J1605 (see Section~\ref{sec:intro}).
Therefore, we conclude that the claimed broad absorption line at $\sim0.8\,$keV in J1605 is most likely model-dependent.

\subsection{An atmospheric origin for the absorption feature}\label{subsec:atmos_abs}

An alternative explanation for the absorption features observed among XINS spectra considers atomic transitions in a hydrogen atmosphere 
\citep{Lai01,Medin06,Kerkwijk07,Medin08,Potekhin14}.
Absorption features resulting from magnetized atmospheres have been considered in, e.g., \citet{Sanwal02,Kerkwijk04,Suleimanov12}.
With a gravitational redshift of z$_g=0.3$ (see Section~\ref{subsec:NSA}), and for a mean value of the absorption feature of $0.44\,$keV, the unredshifted energy of the feature goes up to $0.57\,$keV.
Considering a partially ionized hydrogen atmosphere (see Section~\ref{subsec:NSA}), such a value does not correspond to expected energies except possibly a transition from ground state to first excited tightly bound state ($\nu=0$, $s=0\rightarrow1$, where $\nu$, $s$ are the principal and magnetic quantum numbers, respectively) and only if $B>10^{13}$~G (see e.g., \citealt{Lai01, Medin06,Kerkwijk07,Ho08}).
Such a high magnetic field is also required for the observed feature to be associated with spectral features due to a condensed surface spectrum (see e.g., \citealt{Potekhin12,Hambaryan17}).
However, for a magnetic field strength as high as that implied by the pCF ($9.0\times10^{13}\,$G, see Section~\ref{subsec:pcrsf}), absorption features may be washed out due to the vacuum resonance mode conversion \citep{Lai+Ho03, Ho+Lai03, Potekhin04, vanAdelsberg06,Kerkwijk07, Potekhin12, Potekhin14}.

Even though our best-fit atmosphere models are composed of hydrogen, we notice that in \citet{Medin08} there is an expected bound-free transition at $564.9\,$eV in a helium atmosphere with a magnetic field of $10^{13}\,$G.
This energy is consistent with the unredshifted feature's nominal energy found by the best-fit double-blackbody model, that is $0.565\,$keV.

\subsection{Lack of pulsations}

A periodic X-ray signal associated with the rotation period of J1605 remains undetected. 
Non-detection of pulsation for this source is also reported by \citet{Pires19}. The upper limit on the pulsed fraction of $\sim$1.3\% we have derived is comparable to the very low modulation fraction of 1.2\% observed for another XINS, RX J1856.5$-$3754 \citep{Tiengo07}. Such a low level of pulsations indicates that the variations of the apparent thermal flux caused by the changing view of the surface of the neutron star as it rotates are modest. This could be a consequence of a close alignment between the observer's line of sight and the spin axis of the NS.
However, for the case of double-blackbody emission, \citet{Pires19} argue that the likelihood that we do not see pulsations from the source due to the particularly unfavorable viewing geometry is small, ${\sim}2\%$.


Moreover, or alternatively, the temperature contrast across the stellar surface may be low, or the heat distribution may be approximately symmetric about the rotation axis, which when combined with relativistic light bending (which acts to suppress the amplitude of thermal pulsations) can result in a low pulsed fraction.
At least in the case of the double-blackbody model (see Section~\ref{subsec:bbody}), the temperature contrast between the hot spot and the cooling NS is large,  favoring the symmetrical heat distribution interpretation.

On the other hand, the fit of atmospheric models hints to a uniform temperature across the entire surface of the NS (see Section~\ref{subsec:NSA}), which would lead to isotropic emission, thus to the observed lack of pulsation.
Although the above represents a tempting scenario, we note that an inhomogeneous temperature distribution is expected across the NS surface due to the presence of strong magnetic fields (see \citealt{2007MNRAS.380...71H,Page07,Pons09,Hambaryan17}, and present work). Nonetheless, other theoretical works (e.g., \citealt{Potekhin03, Kaminker06}) do not expect strong deviations of the temperature distribution from spherical symmetry, even in the case of strong magnetic fields, e.g., if a ``patched'' multipole geometry of the magnetic field is present (see \citealt{Perez06}).

\section{Summary}

We have performed X-ray spectral and timing analyses of the XINS RX~J1605.3+3249 observed with \textit{NICER} and \textit{XMM}.
Our main results can be summarized as follows:

\begin{itemize}
\item In agreement with \citet{Pires19}, we found no evidence of pulsation for J1605 with pulsed fraction greater than $1.3\%$ ($3\sigma$) for periods above 0.15 s.
With \textit{NICER}, we find no pulsations with pulsed fraction greater than $2.6\%$ for periods above 2 ms.
As such, J1605 remains the only XINS member that exhibits no pulsation.
This may be due to either geometrical effects or to isotropic atmospheric emission.

\item The X-ray spectrum of J1605 is equally well fitted by a double-blackbody model and by two magnetic atmospheric models: the \texttt{NSA}/\texttt{NSA}-like model and the \texttt{NSMAXG} model.
Double-blackbody and \texttt{NSA}-like models predict consistent flux values at optical wavelengths.
The former also fits optical archival data.
A color-corrected version of the \texttt{NSA} model also fits the optical data.

\item Those models that fit optical data show a UV excess that, if due to absorption by material surrounding the NS, would require a column density of a few times $10^{21}\,$cm$^{-2}$.
However IR archival data are difficult to reconcile with this scenario.
If the best-fit double-blackbody model with Galactic N$_\texttt{H}$ is considered, UV data are also well fitted, while optical data can be accounted as blackbody emission at kT$_\texttt{eff}=1\,$keV from a fallback disk.

\item The X-ray spectrum of J1605 shows an absorption feature at $\sim0.44\,$keV, consistent with prior analyses but without the harmonically related features previously claimed.
Our analysis favors the interpretation of this feature as a proton cyclotron resonant feature, implying a magnetic field strength of $9\times10^{13}\,$G.
Contribution to this feature may come from atomic hydrogen transition from the ground state to the first excited tightly bound state.
\end{itemize}

The lack of a well established value of the distance to J1605 prevents us from inferring additional physical characteristics of this source.
It is therefore of key importance to perform further observations in order to measure J1605's distance conclusively.

\acknowledgments
We thank the referee, Dr. Adriana Pires, for carefully reading of the
manuscript and for the detailed examination of the analyzed data, which helped to improve the paper.

This work was supported in part by NASA through the NICER mission and the Astrophysics Explorers Program. A portion of the results presented were based on observations obtained with \textit{XMM-Newton}, an ESA science mission with instruments and contributions directly funded by ESA Member States and NASA. This research has made use of data and software provided by the High Energy Astrophysics Science Archive Research Center (HEASARC), which is a service of the Astrophysics Science Division at NASA/GSFC and the High Energy Astrophysics Division of the Smithsonian Astrophysical Observatory.  We acknowledge extensive use of the NASA Abstract Database Service (ADS) and the ArXiv.
C.M. is supported by an appointment to the NASA Postdoctoral Program at the Marshall Space Flight Center, administered by Universities Space Research Association under contract with NASA.
\facilities{\textit{NICER}, {\textit{XMM-Newton}}}

\bibliographystyle{yahapj}
\bibliography{references}

\end{document}